\documentclass[a4paper,fleqn]{cas-dc}
\usepackage[authoryear,compress]{natbib}
\usepackage{xurl}
\usepackage{tabularx}
\usepackage{float}

\newcommand{\DO}{\textrm{Oxy}}
\newcommand{\degree}{$^\circ$C}
\newcommand{\Qten}{\ensuremath{Q_{10}}}

\begin{document}
\let\WriteBookmarks\relax
\def\floatpagepagefraction{1}
\def\textpagefraction{.001}

\title [mode = title]{OxyPOM: a biogeochemical model for Oxygen and Particulate Organic Matter dynamics with detailed temperature sensitivity}

\shorttitle{Oxygen and Particulate Organic Matter}
\shortauthors{García-Oliva and Lemmen}

\author{Ovidio García-Oliva}[
                        orcid=0000-0001-6060-2001]
\cormark[1]
\ead{ovidio.garcia@hereon.de}
\credit{Conceptualization, Methodology, Data curation, Formal and statistical analysis, Software, Writing-original draft, Writing--review, and editing}

\affiliation{organization={Helmholtz-Zentrum Hereon},
    addressline={Max-Planck-Straße 1}, 
    city={Geesthacht},
    postcode={21502}, 
    country={Germany}}

\author{Carsten Lemmen}[orcid=0000-0003-3483-6036]
\ead{carsten.lemmen@hereon.de}
\credit{Funding acquisition, Conceptualization, Software, Writing-original draft, Writing--review, and editing}

\begin{abstract}
Periods of low dissolved oxygen concentration--hypoxia and anoxia--threaten the health of aquatic ecosystems and the services they provide.
Hypoxia is strongly influenced by temperature, but the different sensitivities and response functions of oxygen removal and production processes to temperature are not regarded in most models.
Here we present OxyPOM---Oxygen and Particulate Organic Matter, a nuanced temperature-aware process-based biogeochemical model.
OxyPOM incorporates nuanced temperature sensitivities for the key oxygen-related processes photosynthesis, re-aeration, respiration, mineralization, and nitrification.  Further sensitive variables like optimal light intensity, winter grazing inhibition, and pathogenesis are also represented.
Our model was tested in an idealized water column experiment, representing a typical estuarine seasonal low-oxygen environment.  
Differences between nuanced and uniform temperature sensitivities affect seasonal patterns of oxygen-related processes, resulting in under- or overestimation during different times of the year, particularly with higher differences in summer. 
While these changes may balance in the overall annual oxygen budget, uniform sensitivities underestimate particulate organic carbon production by up to a factor of four along the year and overestimate nutrient concentrations.
This nuanced approach to temperature sensitivity allows us to explore and test new hypotheses related to climate warming and heatwaves, addressing the ecosystem changes demanded by climate change models.
\end{abstract}


\begin{highlights}
  \item OxyPOM has nuanced temperature sensitivities
  \item Key processes are photosynthesis, re-aeration, respiration, mineralization, and nitrification
  \item OxyPOM includes optimal light intensity, winter grazing inhibition, and pathogenesis
  \item Budgets of POM and nutrients are sensitive to nuanced temperature responses
\end{highlights}

\begin{keywords}
  Biochemical model \sep
  Temperature sensitivity \sep
  Water quality \sep
  Hypoxia \sep
  Phytoplankton viral infections 
\end{keywords}

\maketitle
\section{Introduction}

Dissolved oxygen is a fundamental indicator of water quality and the health of aquatic ecosystems \citep{EC2006, Kannel2007}. 
Reduced oxygen concentrations, specially hypoxia (below 4\,mg\,L\textsuperscript{-1}) and anoxia (below \,mg\,L\textsuperscript{-1}), adversely affect aerobic organisms and promote the release of greenhouse gases, nutrients, and toxins from sediments \citep{Bastviken2011, Salk2016}. 

The oxygen budget in aquatic systems is governed by biogeochemical (BGC) processes that in model applications act as sources (e.g., photosynthesis) or sinks (e.g., organic matter mineralization, respiration, nitrification, and other biotic and abiotic reactions) of oxygen \citep{Yakushev2013, Holzwarth2018a}. 
In addition, air-water oxygen exchange (re-aeration) functions as a source or sink depending on the temperature-dependent saturation state of the water \citep{Wanninkhof2014, Carter2021}.
These controlling variables exhibit pronounced seasonal variability due to their temperature sensitivity, leading to shifts in the dominant processes throughout the year. 
Substantial changes in oxygenation and deoxygenation can also occur due to  modifications in particulate matter loads and, consequently, water turbidity, light availability, nutrient concentrations, and organic matter levels \citep{Graham2024, Bernal2025}, thereby influencing, for example, photosynthetic rates, and particulate organic matter (POM) production.

Water temperature plays a particularly critical role, as both long-term warming trends and short-term heat waves are closely associated with observed deoxygenation rates \citep{Blaszczak2023}.
Nevertheless, the relative effect of temperature changes in the process leading to oxygen dynamics remain poorly constrained in current model applications.
Most existing water quality and ecosystem models describe oxygen dynamics primarily as a byproduct of more or less complex biotic and abiotic interactions, with only a few focusing explicitly on the processes that directly govern the oxygen budget (Table~\ref{table:models}). 
Moreover, many models that are used to study oxygen dynamics lack the necessary temperature sensitivities to assess the effects of temperature on oxygen responses, either because the relevant processes are absent or because their temperature dependencies are poorly represented or entirely missing. 
Additionally, the recognition of previously neglected processes, such as pathogenesis \citep{Weitz2015, Wirtz2019, Krishna2024}, adds further complexity to current models, limiting their applicability for accurate oxygen modeling.  

Here we present the biogeochemical model OxyPOM (\textbf{Oxy}gen and \textbf{P}articulate \textbf{O}rganic \textbf{M}atter), which include individual temperature sensitivities in oxygen-related processes as well as in other biogeochemical rates of lower trophic levels.
Our objective is to highlight the impact of specific temperature sensitivities in ecosystem dynamics, we compared two cases of temperature sensitivity formulations (1)~all process share a unique sensitivity parameter and (2)~each process have its own parameter. 
We coupled OxyPOM to a one-dimensional, depth-resolving hydrodynamical setup in GOTM to simulate oxygen dynamics in an idealized case, representing a typical estuarine situation with seasonal hypoxia.
Our primary goal is to introduce OxyPOM as a model suitable for the quantification of the relative importance of biogeochemical and physical processes driving oxygen and POM dynamics, with emphasis on the role of temperature sensitivities. 

\begin{table*}[b]
\centering
\caption{Temperature sensitivity terms in OxyPOM and comparison with other ecosystem/oxygen models.
Marked with * are the models used to evaluate oxygen dynamics available in the literature. 
Abbreviations: Aer = Reaeration; Pho = Photosynthesis/Phytoplankton/Primary production; Res = Respiration; Nit = Nitrification; Min = Mineralization;
OLI = Optimal Light Intensity; Het = Heterotrophy/Grazing; VBS = Virus Burst Size; VIn = Virus Inactivation.
Temperature sensitivity: ``+'' = explicit independent term; ``/'' = partial/shared; ``--'' = not included; ``?'' = unclear/not documented. Blank spaces mark when a process is not considered by the model.}
\label{table:models}
\renewcommand{\arraystretch}{1.2}
\begin{tabular}{lccccccccc}
\hline
\textbf{Model} &
\textbf{Aer} &
\textbf{Pho} &
\textbf{Res} &
\textbf{Nit} &
\textbf{Min} &
\textbf{OLI} &
\textbf{Het} &
\textbf{VBS} &
\textbf{VIn} \\
\hline

*Estuary DO model \citep{Holzwarth2018a} &
+ & + & + & + & + & + & + &  &  \\

*ERGOM \citep{Neumann2022} &
+ & / & / & + & + &  & / &  &  \\

*ECOSMO \citep{Schrum2006} &
+ & -- & -- & -- & -- &  & -- &  &  \\

*QSim \citep{Scbol2014} &
+ & + & ? & ? & + &  & ? &  &  \\ 

*BROM \citep{Yakushev2017} &
+ & -- & -- & / & / &  & / &  &  \\ 

*PCLake \citep{Hu2016} &
/ & + & + & + & + & ? & + &  &  \\

AQUATOX \citep{Park2008} &
+ & + & + & + & + &  & + &  &  \\ 

2MPPD model \citep{Maar2011} &
 -- & + &  &  & + &  & + &  &  \\

MAECS \citep{Wirtz2019} &
 & + & + &  & + & / & + & / & + \\

Bio-Vi \citep{Krishna2024} &
 & + &  &   &  +  &  & + & -- & + \\

OxyPOM (this study) &
+ & + & + & + & + & + & + & + & + \\








\hline
\end{tabular}
\end{table*}

\section{Methodology}

\subsection{OxyPOM model description}\label{sm:model-description}
A predecessor version of OxyPOM was initially implemented by \citet{Holzwarth2018a} in the context of the closed-source Unstructured Tidal Residual Model--Delft Water Quality (UnTRIM DELWAQ) system used by the German Federal Waterways Engineering and Research Institute (BAW). 
This implementation lacked many of the processes introduced here, but also faced technical deficiencies regarding portability and FAIR principles \citep[Findable, Accessible, Interoperable, Reusable][]{Wilkinson2016} and Good Modeling Software Practices \citep{Lemmen2024}.
Our implementation as open source utilizes the Framework for Aquatic Biogeochemical Models Application Programming Interface \citep[FABM API,][]{Bruggeman2014}, thereby ensuring accessibility, interoperability with other aquatic process models,  reusability across different hydrodynamic models, and zero- to three-dimensional domains.
Beyond the original implementation, we incorporated vertically-explicit formulations for re-aeration in rivers and estuaries \citep{Raymond2001}, primary production, and light attenuation; for better resolving effects of temperature variations we included additional mortality terms for phytoplankton, accounting for viral infections \citep{Wirtz2019} and temperature-sensitive loss rates \citep{Scharfe2009}.

The temperature sensitivity of biogeochemical rates is formulated as a $\Qten$ rule \citep{vantHoff1884, Mundim2020}:
\begin{equation}
    \tau_{K}(T) = Q_{10,K}^{\frac{T-T^*}{10}}, 
    \label{eq:Q10}
\end{equation}

where $T$ is the temperature, $Q_{10,K}$ is the coefficient for the temperature sensitivity of $K$, which  represents a temperature-sensitive process or variable, and $T^*$ is a reference temperature, in this case 20~\degree.

Dissolved oxygen budget follows a mass balance equation of major processes producing/consuming oxygen:

\begin{equation}\begin{split}
 \frac{\textrm{d} \DO}{\textrm{d}t} & =  \textrm{Re-aeration} + \textrm{Photosynthesis} \\
 & - \textrm{Respiration} - \textrm{Nitrification} - \textrm{Mineralization}.
 \label{eq:DO}\end{split}
\end{equation}

\begin{figure*}[tb] 
 \centering
 \includegraphics[page=1,width=0.99\textwidth]{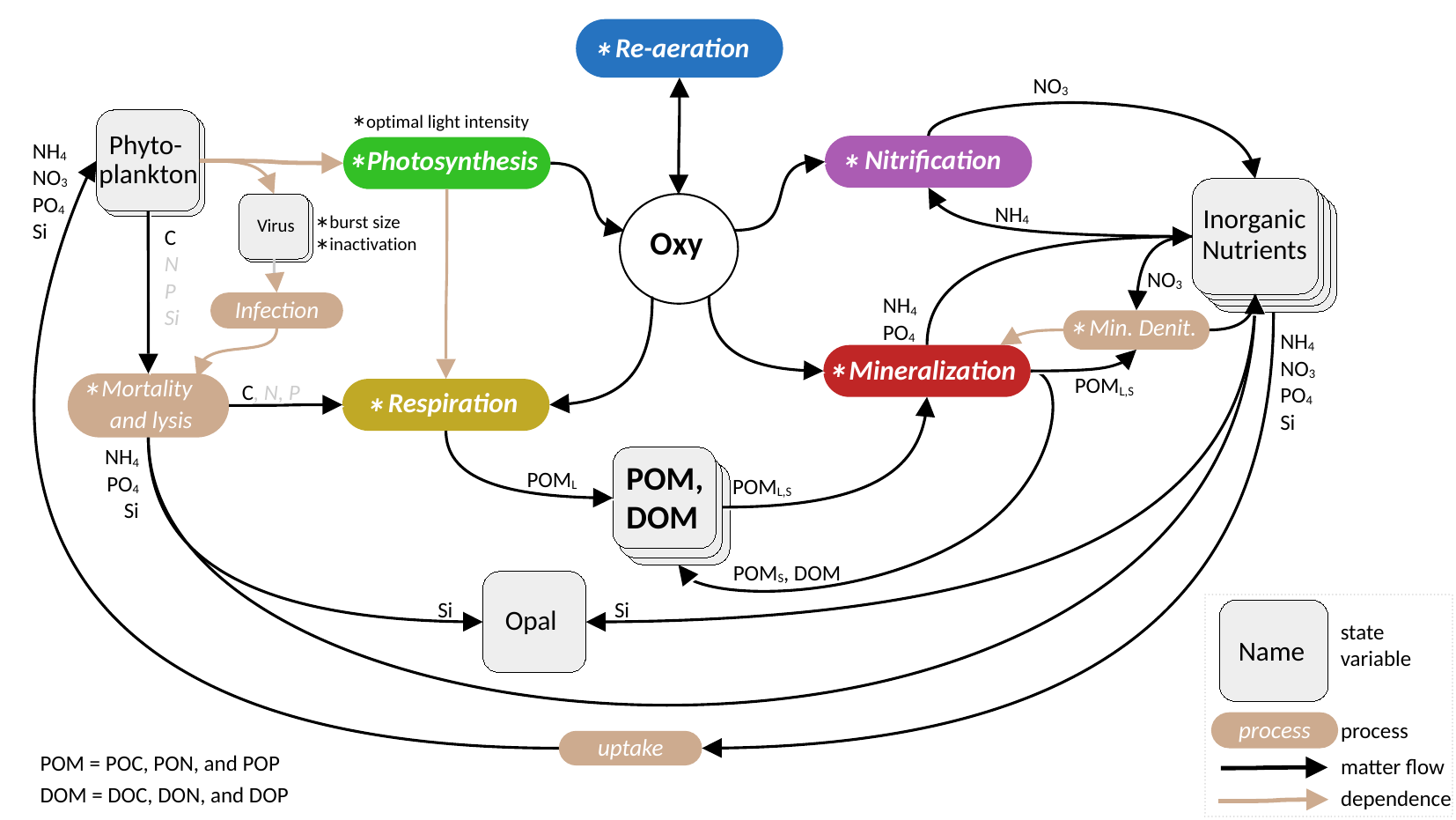} 
 \caption{\textbf{Model configuration:} 
 Processes resolved by OxyPOM, except for settling and light attenuation. 
 Related state variables are shown as grey boxes.  
 Grey symbols denote implicit fluxes defined by a fixed phytoplankton stoichiometry.
 Process can change the nature of a flux, for example POM (explicitly represented by carbon) is converted into inorganic nutrients (ammonia and phosphate) through the process of ``mineralization''.
Processes or variables that include temperature sensitivity are marked with an asterisk.
 Min. Denit. = mineral denitrification, fraction of total mineralized organic material achieved via  denitrification, i.e., the consumption of oxygen in nitrate to oxidize organic matter liberating molecular nitrogen. 
 }
 \label{fig:model}
\end{figure*}

\subsubsection{Re-aeration and water-air oxygen exchange}
Oxygen transport due to aeration is calculated using the phenomenological equations of \citet{Wanninkhof1992,Wanninkhof2014}.
The oxygen surface flux is

\begin{equation}
 \textrm{Re-aeration} = \alpha \cdot k_{\textrm{\scriptsize air}}\cdot(\textrm{Sat} - \DO),
 \label{eq:Aer}
\end{equation}

where $\alpha$ is a form factor that corrects surface to volume ratio of idealized topologies \citep{Holzwarth2018a}, oxygen is the dissolved oxygen in water, and Sat is the saturation oxygen concentration in water as function of temperature and salinity after \citet{Weiss1970}.
 $k_{\textrm{\scriptsize air}}$ is the coefficient of re-aeration, which we defined as a function by parts corrected in low wind speeds after \citet{Raymond2001}

\begin{equation}
 k_{\textrm{\scriptsize air}} = \left(\frac{\textrm{Sc}}{660}\right)^{-0.5} \cdot
 \begin{cases}
 0.0283\cdot w^3, & \textrm{if } w \geq 11\, \textrm{m\,s\textsuperscript{-1}} \\
 0.3120\cdot w^2, & \textrm{if } 11 > w \geq 3\, \textrm{m\,s\textsuperscript{-1}}\\
 0.9836\cdot \textrm{e}^{0.35\cdot w}, & \textrm{if } w < 3\, \textrm{m\,s\textsuperscript{-1}}
 \end{cases}
 \label{eq:krear}
\end{equation}

where Sc is the Schmidt number, which is a quartic function of the temperature and depends on the salinity \citep{Wanninkhof2014}, and $w$ is the wind velocity in m\,s\textsuperscript{-1}.
This formulation is new to the model as the original implementation was a vertical-averaged description \citep{Holzwarth2018a} and here we redefined the process to be vertically explicit.

\subsubsection{Phytoplankton}
The two phytoplankton (micro-algae) classes--one with dependence on dissolved silicate, thus representing diatoms--have growth rates that depend on photosynthesis, which is limited by light availability, and dissolved silica, nitrogen and ortho-phosphate concentrations.
 
\begin{equation}
 \frac{\textrm{d} \textrm{Phy}_i}{\textrm{d}t} =  \textrm{Phy}_i \cdot \left( \rho_i - m_i - r_i - \lambda_i \right) ,
 \label{eq:Phy}
\end{equation}

where $\rho_i$ is the gross production rate, $m_i$ is loss rate (mortality), $r_i$ is the respiration rate and $\lambda_i$ is the virus-related mortality for each phytoplankton.

The gross primary production is defined as

\begin{equation}
 \rho_i = f_{\textrm{\scriptsize light},i}\cdot f_{\textrm{\scriptsize nut},i}\cdot\tau_{\textrm{\scriptsize Phy},i}(T)\cdot \mu^*_i,
 \label{eq:rho}
\end{equation}

where $f_{\textrm{\scriptsize light},i}$ and $f_{\textrm{\scriptsize nut},i}$ are growth limitation coefficients for nutrient and light, $\mu^*_i$ is the maximum growth rate in optimal conditions---neither light nor nutrient limitation---at the reference temperature, and $\tau_{\textrm{\scriptsize Phy},i}(T)$ is the temperature sensitivity terms as defined in Eq.~\ref{eq:Q10} \citep{Montagnes2003}.

The limitation coefficients for light are defined as a saturating exponential 

\begin{equation}
 f_{\textrm{\scriptsize light},i} = 1 - \textrm{e}^{-I/I'_i}, 
 \label{eq:flight}
\end{equation}

where $I$ is the light intensity and $I'_i$ is a reference light intensity.
The light sensitivity is temperature-dependent that follows a  $\Qten$ rule Eq.~\ref{eq:Q10}.
The reference light intensity shows temperature dependence as in \citet{Holzwarth2018a} via 

\begin{equation}
 I'_i=I_i^*\cdot\tau_{I^*}(T), 
 \label{eq:light-temp}
\end{equation}

where $I_i^*$ is the optimal light intensity for each phytoplankton group at the reference temperature.

The limitation coefficients for nutrients are defined as a Liebig's rule of the minimum

\begin{equation}
 f_{\textrm{\scriptsize nut},i} = \min\left(f_{N,i},f_{P,i},f_{Si,i}\right),
 \label{eq:fnut}
\end{equation}

where each $f_X$ follows a Monod equation with half saturation $K_X$, thus

\begin{equation}
  f_{X} = \frac{X}{X+K_X}.
 \label{eq:fnutX}
\end{equation}

As by convention Phy$_1$ represents diatoms---silicate limited---and Phy$_2$ other non-diatom phytoplankton, we set $f_{\text{Si},2} = 1$ to avoid silica limitation in non-diatoms. 

The mortality rate $m_i$ is a piece-wise function, which emulates temperature dependent grazing with inhibition in a factor $m_\text{winter}$  during colder temperatures \citep{Scharfe2009} as

\begin{equation}
 m_i = m^*_i \cdot
 \begin{cases}
 \tau_m(T), & \textrm{for } T \geq 20\,^\circ\textrm{C} \\
 1, & \textrm{for } 5 < T < 20\,^\circ\textrm{C} \\
 m_\text{winter}, & \textrm{for } T \leq 5\,^\circ\textrm{C}.
 \end{cases}
 \label{eq:m}
\end{equation}

The respiration rate $r_i$ is divided in basal and maintenance respiration rates, weighted by a factor $\pi$ as

\begin{equation}
 r_i = \pi\cdot\rho + (1-\pi)\cdot r^*_i\cdot\tau_{\textrm{\scriptsize Res}}(T),
 \label{eq:r}
\end{equation}

where $r^*_i$ is the respiration rate at the reference temperature and $\tau_{\textrm{\scriptsize Res}}$ is the temperature sensitivity.

The virus related mortality or cell lysis $\lambda_i$ depends on a step-function \citep{Wirtz2019} defined as

\begin{equation}
 \lambda_i = \frac{1}{1+\textrm{e}^{S\cdot\left(1-\textrm{\footnotesize Vir}_i\right)}},
 \label{eq:sigma}
\end{equation}

where $\textrm{Vir}_i$ is the extent of the viral infection, and $S$ is a sensitivity of mortality to viral infection.

The oxygen produced by photosynthesis used in the oxygen budget Eq.~\ref{eq:DO} is the addition of the gross primary production of both phytoplankton classes

\begin{equation}
 \textrm{Photosynthesis} = \sum_{i \in (1,2)} \rho_i\cdot\textrm{Phy}_i ,
 \label{eq:Pho}
\end{equation}

and the respiration rate is the sum of the respiration rates plus a rather small fraction $c$ of the mortality rate to emulate grazers respiration

\begin{equation}
 \textrm{Respiration} = \sum_{i \in (1,2)} (r_i + c\cdot m_i )\cdot\textrm{Phy}_i.
 \label{eq:Res}
\end{equation}

\subsubsection{Nutrient uptake}
Nutrient uptake follows a fixed stoichiometry approach where the cell composition remains constant.
The uptake rate is thus proportional to the net primary production, i.e., gross primary production minus respiration.
In the case of phosphorus and silica, which are represented as a single inorganic form, the uptake rates are straightforwardly calculated as

\begin{equation}
 u_X = a_X\cdot\sum_{i \in (1,2)} (\rho_i-r_i)\cdot\textrm{Phy}_i, 
 \label{eq:u1}
\end{equation}

$a_X$ is the stoichiometric ratio of $X$ relative to carbon.

For nitrogen, the ammonia uptake is preferred in a flexible way unless the ammonia concentration is too low when nitrate uptake are instead preferred.
The fraction of ammonia uptake when ammonia concentration is below a critical value of 0.7\,mmol\,m\textsuperscript{-3} is

\begin{equation}
 f_{\textrm{\scriptsize NH}_4} = \frac{\textrm{\small NH}_4}{\textrm{\small NH}_4 + \textrm{\small NO}_3}, 
 \label{eq:fNH4}
\end{equation}

Ammonia and nitrate uptake is thus defined as 

\begin{eqnarray}
 u_{ \textrm{NH}_4} &=& a_N\cdot f_{\textrm{\scriptsize NH}_4}\cdot\sum_{i \in (1,2)} (\rho_i-r_i)\cdot\textrm{Phy}_i, \label{eq:uNH4} \\
 u_{\textrm{NO}_3} &=& a_N\cdot (1-f_{\textrm{\scriptsize NH}_4})\cdot\sum_{i \in (1,2)} (\rho_i-r_i)\cdot\textrm{Phy}_i.
 \label{eq:uNO3}
\end{eqnarray}

In the case of ammonia above the critical value, we assume that nitrate uptake is inhibited, thus setting $f_{\textrm{\scriptsize NH}_4} = 1$.

\subsubsection{Particulate and dissolved organic matter}
POM and DOM have an explicit elemental composition (carbon, nitrogen and phosphorus).
POM is present in two qualities, which transition in the sequences labile $\rightarrow$ semi-labile $\rightarrow$ dissolved and labile $\rightarrow$ dissolved.
Labile and semi-labile POM follow the dynamics:

\begin{eqnarray}
 \frac{\textrm{d} \textrm{PO}X_L}{\textrm{d}t} &=& a_X\cdot(1 - f)\cdot L - d_{X,L\rightarrow S} \nonumber \\ 
  & & - d_{X,L\rightarrow D} - M_{X,L},\label{eq:POXL}\\
 \frac{\textrm{d} \textrm{PO}X_S}{\textrm{d}t} &=& d_{X,L\rightarrow S } - d_{C,S\rightarrow D} - M_{X,S},\label{eq:POXS}
\end{eqnarray}

where $X$ is either carbon, nitrogen or phosphorus, the subindices $L$, $S$, and $D$ are for labile, semi-labile, and dissolved, respectively, $f$ is the fraction of nutrient released by autolysis,  $L = m_1 + m_2 +\lambda_1 +\lambda_2$ is the total loss rate including mortality and lysis rates for both phytoplankton groups, $d_{X,i\rightarrow j}$ are degradation rates from the quality $i$ to the quality $j$, and $M_{X,i}$ are mineralization rates.

POM and DOM mineralize to DOC, nitrogen and ortho-phosphate. 
Unlike for dissolved nutrients, dissolved carbon is calculated using a simple mass balance

\begin{equation}
 \frac{\textrm{d} \textrm{DOC}}{\textrm{d}t} = d_{C,L\rightarrow D} + d_{C,S\rightarrow D} - M_{C,D}.
 \label{eq:DOC}
\end{equation}

POM mineralization rates for the element $X$ of quality $i$ are defined as 
\begin{equation}
 M_{X,i} = k_{\textrm{\scriptsize Min},i}\cdot \tau_{\textrm{\scriptsize Min}}(T) \cdot \textrm{PO}X_i,
 \label{eq:MX}
\end{equation}

where $k_{\textrm{\scriptsize Min},i}$ are mineralization rate constants for the quality $i$ and $\tau_{\textrm{\scriptsize Min}}(T)$ is the temperature sensitivity.

Degradation rates are defined from the mineralization rates as 
\begin{equation}
 d_{X,i\rightarrow j} = \kappa_{i\rightarrow j} \cdot M_{X,i},
 \label{eq:dX}
\end{equation}

where $\kappa_{i\rightarrow j}$ is a factor for the decomposition from the quality $i$ to the quality $j$.

\subsubsection{Dissolved nitrogen and nitrification}

Dissolved inorganic nitrogen is the sum of ammonium and nitrate and ammonium transitions to nitrate as a function of oxygen. 
The mass balance for dissolved inorganic nitrogen is thus

\begin{eqnarray}
 \frac{\textrm{d} \textrm{NH}_4}{\textrm{d}t} &=& a_N\cdot f\cdot L + M_{N} - u_{\textrm{\scriptsize NH}_4}- \gamma, \label{eq:NH4} \\
 \frac{\textrm{d} \textrm{NO}_3}{\textrm{d}t} &=& \gamma - M_{\textrm{\scriptsize NO}_3} - u_{\textrm{\scriptsize NO}_3},
 \label{eq:NO3}
\end{eqnarray}

where $\gamma$ is the ammonia consumed in the nitrification process, $M_{N}$ is the total nitrogen mineralized from POM and DOM ($= \sum_{i \in (L,S,D)} M_{N,i}$), $M_{\textrm{\scriptsize NO}_3}$ is mineral denitrification into N\textsubscript{2}, $u_{\textrm{\scriptsize NH}_4}$ and $u_{\textrm{\scriptsize NO}_3}$ are total phytoplankton uptake rates.

The ammonia used during nitrification depends on ammonia and oxygen concentrations as 

\begin{equation}
 \gamma = k_{\textrm{\scriptsize Nit}}\cdot \tau_{\textrm{\scriptsize Nit}}(T)\cdot \frac{\textrm{\small NH}_4}{\textrm{\small NH}_4 + K_{\textrm{\scriptsize NH}_4}}\cdot \frac{\textrm{\small\DO}}{\textrm{\small\DO} + K^*_{\textrm{\scriptsize\DO}}} ,
 \label{eq:gamma}
\end{equation}

where $K_{\textrm{\scriptsize NH4}_4}$ and $K^*_{\textrm{\scriptsize\DO}}$ are half saturation constants for ammonia and oxygen in nitrification, $k_{\textrm{\scriptsize Nit}}$ is the nitrification rate at the reference temperature and $\tau_{\textrm{\scriptsize Nit}}(T)$ is the temperature sensitivity.

As the nitrification process consumes two moles of oxygen per each mol of nitrified ammonia in a two-step process

\begin{eqnarray*}
 \mathrm{NH_4^+ + 1.5\,O_2} &\rightarrow& \mathrm{NO_2^- + H_2O + 2H^+}\\
 \mathrm{NO_2^- + 0.5\,O_2} &\rightarrow& \mathrm{NO_3^-},
\end{eqnarray*}

the total oxygen consumed during mineralization used in the oxygen budget (Eq.~\ref{eq:DO}) is

\begin{equation}
 \textrm{Nitrification} = 2\cdot\gamma.
 \label{eq:Nit}
\end{equation}

\subsubsection{Mineral denitrification and mineralization}
Mineral denitrification is a fraction of total mineralized organics that is achieved by nitrate denitrification $\Phi_N$ in contrast to oxygen consumption as a competing mechanism---which is proportional to ($1-\Phi_N$)--- thus

\begin{equation}
 M_{\textrm{\scriptsize NO}_3} = \Phi_N \cdot\sum_{i \in (L,S,D)} M_{C,i} \,\, \textrm{with}\,\, \Phi_N = \frac{\phi_N}{\phi_N+\phi_{\textrm{\scriptsize\DO}}}
 \label{eq:MNO3}
\end{equation}

where $\phi_N$ and $\phi_{\textrm{\scriptsize\DO}}$ are contributions of nitrate and oxygen in mineralization defined as

\begin{eqnarray}
 \phi_N &=& \frac{\textrm{\small NO}_3}{\textrm{\small NO}_3 + K_{\textrm{\scriptsize NO}_3}}\cdot \left(1-\frac{\textrm{\small\DO}}{\textrm{\small\DO} + K_{\textrm{\scriptsize\DO}}} \right)\cdot\tau_N(T)
 \label{eq:phiN}\\
 \phi_{\textrm{\scriptsize\DO}} &=& \frac{\textrm{\small\DO}}{\textrm{\small\DO} + K'_{\textrm{\scriptsize\DO}}}\cdot\tau_{\textrm{\scriptsize\DO}}(T), 
 \label{eq:phiDO}
\end{eqnarray}

where $K_{\textrm{\scriptsize NO}_3}$, $K_{\textrm{\scriptsize\DO}}$ and $K'_{\textrm{\scriptsize\DO}}$ are half saturation constants for nitrate in denitrification, oxygen inhibition in denitrification and oxygen consumption in mineralization, and $\tau_i$ are temperature-sensitivities with an appropriate $\Qten$ value.

The oxygen consumed during mineralization used in the oxygen budget (Eq.~\ref{eq:DO}) is

\begin{equation}
 \textrm{Mineralization} = (1 - \Phi_N) \cdot\sum_{i \in (L,S,D)} M_{C,i}.
 \label{eq:Min}
\end{equation}

\subsubsection{Ortho-phosphate and silicate}
Ortho-phosphate is similar to NO\textsubscript{3} and NH\textsubscript{4} dynamics (Eqs.~\ref{eq:NO3} and \ref{eq:NH4}) 

\begin{equation}
 \frac{\textrm{d} \textrm{PO}_4}{\textrm{d}t} = a_P\cdot f\cdot L + M_{P} - u_{\textrm{\scriptsize PO}_4}
 \label{eq:PO4}
\end{equation}

where $M_{P}$ is the total phosphorus mineralized from POM and DOM ($= \sum_{i \in (L,S,D)} M_{P,i}$), and $u_{\textrm{\scriptsize PO4}_4}$ is the total phytoplankton uptake rates for ortho-phosphate.

Unlike nitrogen and phosphorus, silicate is present in dissolved --bio-available-- and particulate mineral (Opal) forms

\begin{eqnarray}
 \frac{\textrm{d} \textrm{Si}} {\textrm{d}t} &=& a_{\textrm{\scriptsize Si}}\cdot f\cdot (m_1+\lambda_1) + D_{\textrm{\scriptsize Si}} - u_{\textrm{\scriptsize Si}} \label{eq:Si},\\
 \frac{\textrm{d} \textrm{Opal}} {\textrm{d}t} &=& a_{\textrm{\scriptsize Si}}\cdot (1-f)\cdot m_1 - D_{\textrm{\scriptsize Si}} 
\end{eqnarray}

where $D_{Si}$ is the dissolution rate of opal to bio-available dissolved silicate defined as

\begin{equation}
 D_{Si} = k_{\text{Si}}\cdot\textrm{Opal}\cdot(\textrm{Si}^*-\textrm{Si}),
 \label{eq:DSi}
\end{equation}

where $k_{Si}$ is opal dissolution reaction rate constant, and $\textrm{Si}^*$ is a reference (saturation) bio-available silicate concentration.
Note that the equations for silicate balance include only one phytoplankton class, which represents diatoms. 

\subsubsection{Pathogenesis--Viral infection}
We consider pathogenesis using two different virus, each one exclusively infecting either Phy$_1$ or Phy$_2$. 
The dynamics in intracellular viral density Vir$_i$ is here described by 

\begin{equation}
 \frac{\textrm{d} \textrm{Vir}_i}{\textrm{d}t} = G_i\cdot n_i - H_i - B_i,
 \label{eq:Vir}
\end{equation}

where $G_i$ is infection-replication, $n_i$ is the burst size (relative number of viral particles produced per host), $H_i$ virus removal by host mortality (lysis), and $B_i$ is virus inactivation.

The infection-replication is expressed as the probability of an infected host contacting a susceptible one. 
This is hindered by high concentrations immune and inert particles. 
Based on collision dynamics, we use 

\begin{equation}
 G_i= G^*\cdot \frac{\textrm{\small Vir}_i^2}{\textrm{\small POC}_L+\textrm{\small POC}_S+\textrm{\small Phy}_j},
 \label{eq:G}
\end{equation}

where $G^*$ is a reference infection rate.
The burst size is a temperature-dependent step-function as

\begin{equation}
 n_i= n^*\cdot \tau_n(T)\cdot\frac{1}{1+\textrm{e}^{S\cdot\left(\textrm{\footnotesize Vir}^*-\textrm{\footnotesize Vir}_i\right)}},
 \label{eq:n}
\end{equation}

where $n^*$ is the reference burst size when Vir$_i$ is large enough, $\tau_n$ is the temperature sensitivity, and Vir$^*_i$ is a half-saturation value.

Virus removal by host mortality is defined as the preferential loss of heavily infected hosts. 
This is expressed as an increased host mortality because infected phytoplankton frequently undergo apoptosis. 
This selective removal of infected individuals leads to a relative increase in the survival of healthy or less infected hosts or species, thereby reducing the average viral density.
This is expressed using trait dynamics formulation \citep{Wirtz1996}

\begin{equation}
 H_i= H_i^* \cdot \textrm{Vir}_i \cdot (\textrm{Vir}^* -\textrm{Vir}_i)\cdot \textrm{e}^{-\frac{\textrm{\tiny Phy}_i}{C} }\cdot\frac{\textrm{\small Vir}_i}{\textrm{\small Vir}_i+\textrm{\small Vir}'} \cdot\lambda_i^2\cdot S\cdot \textrm{e}^{S\cdot(1-\textrm{\footnotesize Vir}_i)},
 \label{eq:H}
\end{equation}

where $H_i^*$ is a host-specific reference defense parameter, Vir$'$ is a reference low-value for virus abundance and $C$ is a constant that modulates plankton diversity \citep{Wirtz2019}.
The first two terms $\textrm{Vir}_i \cdot (\textrm{Vir}^*_i -\textrm{Vir}_i)$ are diversity of the infection levels as calculated as in a binomial trait \citep{Wirtz1996}, and the exponential term $\textrm{e}^{-\frac{\textrm{\tiny Phy}_i}{C} }$ emulates the genetic diversity---coupled to virus susceptibility---reduction during blooming phases.
The virus removal due to preferential loss thus depends on the diversity of the levels of infection host and of the host defense.
The term $\frac{\textrm{\small Vir}_i}{\textrm{\small Vir}_i+\textrm{\small Vir}'}$ mantains the levels of infection to viable levels \citep{Wirtz2019}.
The remaining terms are the first partial derivative of $\lambda_i$ in respect to $\textrm{Vir}_i$, standard method for trait dynamics $\frac{\partial \lambda_i}{\partial \textrm{Vir}_i} = \lambda_i^2\cdot S\cdot \textrm{e}^{S\cdot(1-\textrm{\footnotesize Vir}_i)}$

Virus inactivation $B_i$ is expressed by

\begin{equation}
 B_i= B^*\cdot \tau_B(T)\cdot\frac{\textrm{\small Vir}_i^2}{\textrm{\small Vir}_i+\textrm{\small Vir}'},
 \label{eq:B}
\end{equation}

where $B^*$ is a reference inactivation value, and $\tau_B$ is the temperature dependence.

\subsubsection{Light attenuation}
Light attenuation follows an exponential decay with an exponential coefficient $\zeta$ defined by

\begin{equation}
  \zeta = \zeta_0 + \epsilon_\textrm{POC}\cdot\textrm{POC} + \epsilon_\textrm{Phy}\cdot\textrm{Phy} + \epsilon_\textrm{ISPM}\cdot \textrm{ISPM},
  \label{eq:zeta}
\end{equation}

where $\textrm{POC} = \textrm{POC}_L+\textrm{POC}_S$ is the total POC, $\textrm{Phy} = \textrm{Phy}_1+\textrm{Phy}_2$ is the total phytoplankton concentration, $\zeta_0$ is the background attenuation coefficient, $\epsilon_\textrm{POC}$, $\epsilon_\textrm{Phy}$ and $\epsilon_\textrm{ISPM}$ are specific attenuation coefficients for POM, phytoplankton, and a constant concentration of inorganic suspended particulate matter (ISPM), respectively.

\subsubsection{Oxygen flux}
We calculate and compare the relative importance of each oxygen-related process using the oxygen flux $\mathcal{F}$ (i.e., oxygen transported/consumed/produced per unit or area).
The oxygen flux $\mathcal{F}$ is formally defined as the vertically integrated oxygen dynamics caused by each process (Eq.~\ref{eq:DO}) from the surface ($\zeta=0$) to the maximum depth $\zeta_{\max}$.
For example, the oxygen flux due to the process $K$ appearing in Eq.~\ref{eq:DO} is defined by 

\begin{equation}
    \mathcal{F}_{K} = \int_{\zeta=0}^{\zeta_{\max}}K(\zeta) \cdot \textrm{d}\zeta.
    \label{eq:flux}
\end{equation}

\subsection{\emph{In silico} experiments}

\begin{table*}[b]
\centering
\caption{$\Qten$ coefficients for the temperature sensitivities used in the reference case. The reported ranges comprise the maximum and minimum reported values. $^\dag$: A range for virus burst size has not been reported and the displayed values are estimations.}
\label{table:q10}

\begin{tabular}{lllll}
\hline
Parameter name & Symbol & Value & Range & Reference\\

\hline 
phytoplankton growth rate & $Q_{10,\text{Phy}}$ & 1.20 & 1--7 & \citet{Robinson1993} \\  
mortality rate due to grazing & $Q_{10,m}$ & 4.40 & 2.2--4.8 & \citet{Verity2002, Maar2011} \\ 
respiration & $Q_{10,\text{Res}}$ & 1.10 & 1.1--5.6 & \citet{Robinson1993, Ferreira2022} \\ 
mineralization & $Q_{10,\text{Min}}$ & 1.20 & 1--5 & \citet{Hansen2014, Johannsson2025}\\ 
nitrification & $Q_{10,\text{Nit}}$ & 1.12 & 0.2--2.9 & \citet{Krishnan2014}\\ 
optimal light intensity & $Q_{10,I^*}$ & 1.04 & 1.5--1.9 & \citet{Edwards2016,Coles2000}\\
virus burst size$^\dag$ & $Q_{10,n}$ & 4.40 & $<1$--4.4 & \citet{Padhy1977,Nagasaki2003}\\ 
virus inactivation & $Q_{10,B}$ & 1.48 & 1.3--2.7 & \citet{Yap2021a}\\ 

\hline
\end{tabular}
\end{table*}

We use an idealized 20-m depth, vertically-resolved, one-dimensional Generalized Ocean Turbulence Model (1D-GOTM) setup inspired by the estuary scenario (available from \url{https://gotm.net/cases/estuary}), which is similar to the one used in several published applications \citep[e.g.,][]{Burchard2010}.
The parameterization for OxyPOM was derived for the Elbe estuary \citep{Holzwarth2018a, Garcia-Oliva2025a} (Table~\ref{table:parameters}) and the meteorological forcing (temperature and wind speed) was inspired by the climatology of this region (53$^\circ$52' N, 8$^\circ$42' W).

In this setup, we compare two OxyPOM parameterizations.
The first using uniform temperature sensitivities, and the second with specific temperature sensitivities ($\Qten$ values specific to each process, Table~\ref{table:q10}) and used as the reference case.
For the case with uniform sensitivity, all $\Qten$ coefficients were set to their geometric mean ($=2.5$)
We simulated the system for two consecutive years and to avoid artifacts derived from the initial conditions, only the second year was used in the data analyses. 

The temporal and spatial differences between the reference and the uniform cases were evaluated using the difference between the uniform and reference cases $x_{\text{uni}}-x_{\text{ref}}$, where $x_.$ represents each variable.

We also evaluated the percentage relative difference $E$, which compares how much two values differ relative to their average magnitude:

\begin{equation}
 E = 50\cdot\frac{\bar{x}_{\text{uni}}-\bar{x}_{\text{ref}}}{\bar{x}_{\text{uni}}+\bar{x}_{\text{ref}}}
 \label{eq:E},
\end{equation}

where $\bar{x}_{\text{uni}}$ and $\bar{x}_{\text{ref}}$ are the daily vertically-averaged results for each variable in the uniform temperature sensitivity and reference cases, respectively.

Finally, we evaluate the relation between temperature $T$ and percentage relative difference $E$ using the 2D density plot of these variables.
All data analyses and figures were prepared in R \citep{RCoreTeam2024}, using the packages ncdf4, lubridate, TeachingDemos, MASS, scales, and latex2exp.

\section{Results}

\begin{figure*}[htb] 
 \centering
 \includegraphics[page=1,width=.90\textwidth]{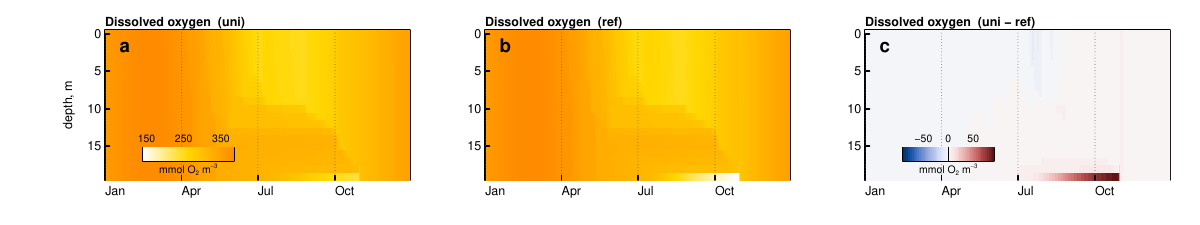} 
 \includegraphics[page=3,width=.90\textwidth]{figs/comparison} 
 \includegraphics[page=5,width=.90\textwidth]{figs/comparison} 
 \includegraphics[page=7,width=.90\textwidth]{figs/comparison} 
 \includegraphics[page=9,width=.90\textwidth]{figs/comparison} 
 \includegraphics[page=11,width=.90\textwidth]{figs/comparison} 
 \includegraphics[page=13,width=.90\textwidth]{figs/comparison} 
 \caption{\textbf{Spatial and temporal dynamics of key ecological variables in uniform-sensitive (uni) and reference (ref) cases, and their difference (left, center, and right columns, respectively).} 
 Dissolved oxygen (a--c), total phytoplankton (d--f), DOC (g--i), total POC (j--l), ammonia (m--o), nitrate (p--r), and phosphate (s--u).
 }
 \label{fig:main_vars}
\end{figure*}

\begin{figure*}[htb] 
 \centering
\includegraphics[page=16,width=.90\textwidth]{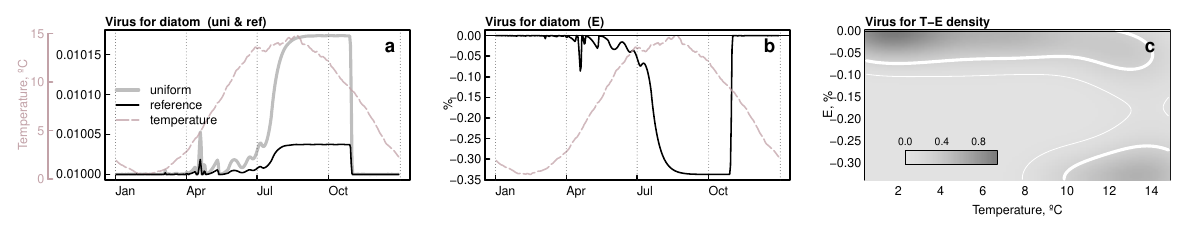} 
 \includegraphics[page=18,width=.90\textwidth]{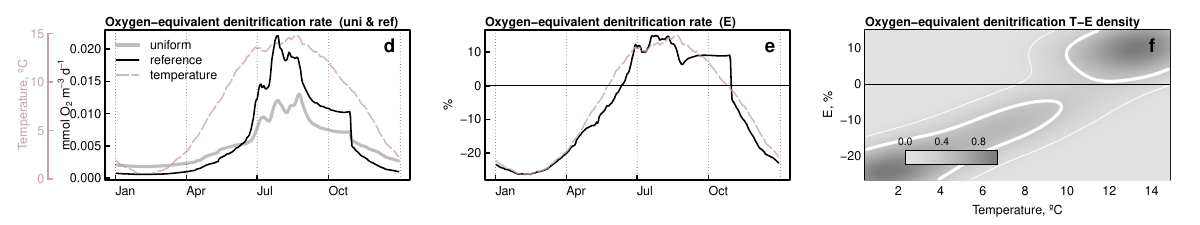} 
 \includegraphics[page=20,width=.90\textwidth]{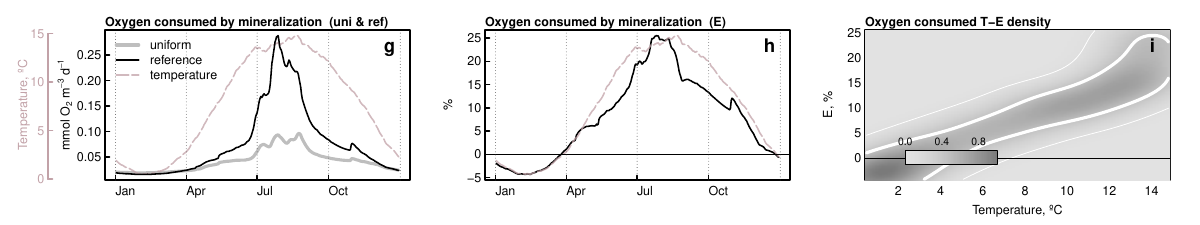} 
 \includegraphics[page=22,width=.90\textwidth]{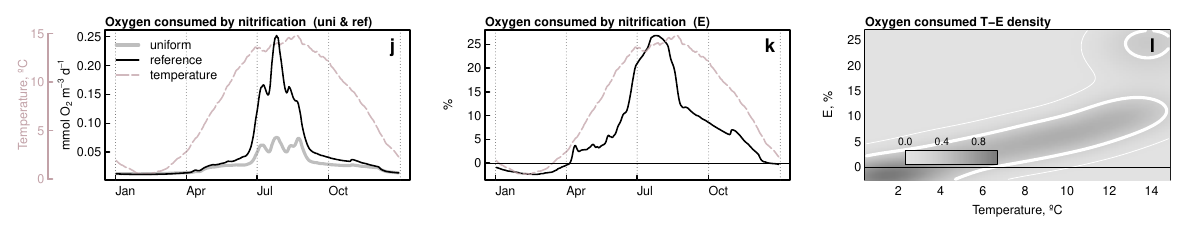} 
 \includegraphics[page=24,width=.90\textwidth]{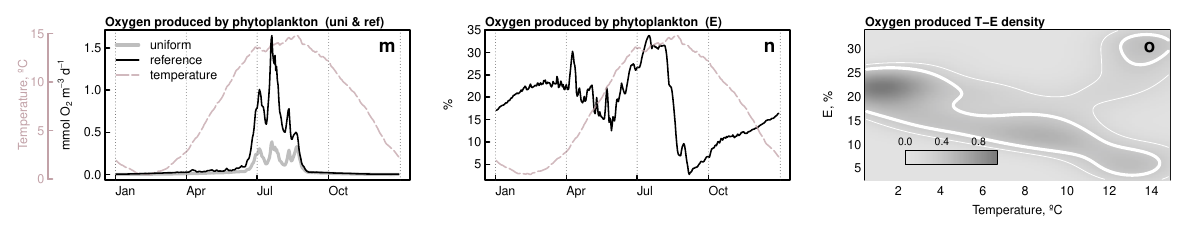} 
 \includegraphics[page=26,width=.90\textwidth]{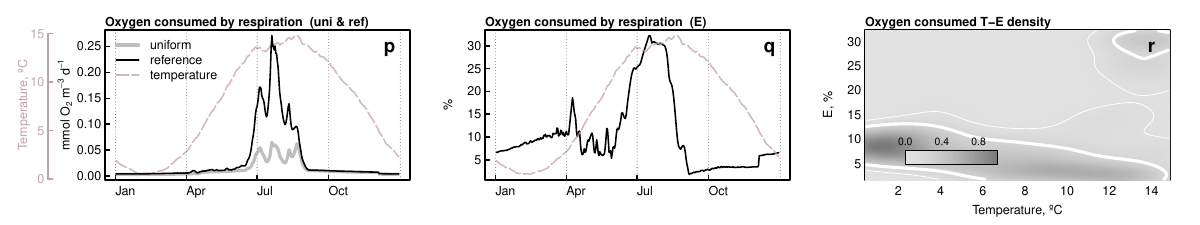} 
 \caption{\textbf{Vertical-averaged virus and oxygen-related process in uniform-sensitive (uni) and reference (ref) cases, percent relative difference $E$, and $T$-$E$ density plot (left, center, and right columns, respectively).} 
 Virus for diatom (a-c), denitrification (d-f), oxygen consumed by mineralization (g-i) and nitrification (j-l), produced by phytoplankton (m-o), and consumed by respiration (p-r).
 The pink dashed line shows the temperature.
 In the density plots (right column), white lines show the contour containing 75\% and 50\% of the values (thin and thick lines, respectively).
 }
 \label{fig:oxygen_related}
\end{figure*}

\subsection{Ecological variables}

Dissolved oxygen, total phytoplankton, DOC, total POC, and nutrient concentrations show important differences in their spatial and temporal distributions depending on the assumed temperature sensitivities  (Fig.~\ref{fig:main_vars}, see the vertical-averaged values in Fig.~\ref{fig:main_vars2}).
Dissolved oxygen concentrations are similar in both cases (i.e., uniform-sensitivity and reference) with an overestimation of 50--100 mmol O\textsubscript{2} m\textsuperscript{-3} in Jul--Oct in the bottom layer in the uniform-sensitivity case (Figs.~\ref{fig:main_vars}a-c).
The reference case shows more pronounced phytoplankton blooms in summer than the uniform case (Figs.~\ref{fig:main_vars}d and e).
Total phytoplankton is underestimated in the top 10 m in Jul-Aug by 5--10 mmol C m\textsuperscript{-3} (Figs.~\ref{fig:main_vars}f).
DOC is slightly underestimated in the uniform case in all seasons and depths, with DOC accumulation in the bottom in Jul-Oct with values 10--40 mmol C m\textsuperscript{-3} (Figs.~\ref{fig:main_vars}g-i).
Total POC is understimated in the uniform case; higher overestimation starts below 15 m in Apr-May, up to Oct in the bottom layer with values 100--200 mmol C m\textsuperscript{-3} (Figs.~\ref{fig:main_vars}j-l). 

Contrary to carbon-related variables (i.e., total phytoplankton, DOC, and total POC), concentrations of ammonia, nitrate, and phosphate are overestimated in the uniform sensitivity case most of the time (Figs.~\ref{fig:main_vars}m-u).
Ammonia is slightly overestimated by approximately 0.1 mmol N m\textsuperscript{-3}, except for high underestimation during the phytoplankton bloom in the top 10 m and in the bottom layer  (Figs.~\ref{fig:main_vars}o).
Nitrate is overestimated in the uniform case before the phytoplankton bloom by approximately 2 mmol N m\textsuperscript{-3}. It increases above 5 mmol N m\textsuperscript{-3} after the bloom (Figs.~\ref{fig:main_vars}r).
Phosphate shows a pattern similar to nitrate (Figs.~\ref{fig:main_vars}s-u):  before the phytoplankton bloom, phosphate is higher by approximately 0.25 mmol P m\textsuperscript{-3} and increases above 0.5 mmol N m\textsuperscript{-3} after the bloom in the uniform case (Figs.~\ref{fig:main_vars}u).
Ammonia, nitrate, and phosphate, as well as carbon-related variables, are underestimated in the bottom layer in Jul-Oct, the same time and place where oxygen is overestimated (right column in Figs.~\ref{fig:main_vars}).

\subsection{Temperature sensitivity: virus and oxygen-related processes}
The vertical-averaged virus and oxygen-related process shows a strong seasonal dynamics, which depends on the assumed temperature sensitivity and is strongly linked to temperature (Fig.~\ref{fig:oxygen_related}, see the spatial and temporal dynamics in Fig.~\ref{fig:oxygen_related2}).
Virus shows only small differences between the uniform and reference temperature sensitivity, with larger differences observed in Jul-Oct and some peaks in Apr; these differences do not closely follow temperature (Fig.~\ref{fig:oxygen_related}a and b).
The density plot of percent relative difference and temperature shows that higher absolute differences occur between 10 and 14 \degree\ but are close to 0 in all other temperatures (Fig.~\ref{fig:oxygen_related}c). 

Denitrification rate follows temperature changes, with higher rates in summer (Fig.~\ref{fig:oxygen_related}d), while the relative difference ranges from -25 to 15\% that closely follows temperature changes with exceptions in Oct,  denitrification is lower during the entire year but summer months (Fig.~\ref{fig:oxygen_related}e and d).
The oxygen consumed by mineralization follows temperature changes, with higher rates in summer (Fig.~\ref{fig:oxygen_related}g), while the relative difference ranging from -5 to 25\% closely follows temperature changes with exceptions in Aug-Oct (Fig.~\ref{fig:oxygen_related}h and i).
The oxygen consumed by nitrification follows temperature changes, with higher rates in summer and strong increases in Jul (Fig.~\ref{fig:oxygen_related}g), while the relative difference ranging from -5 to 25\%  loosely follows temperature changes(Fig.~\ref{fig:oxygen_related}k), for example, showing higher differences at 14 \degree\ than the expected by the trend observed at lower temperatures (high-density spot in the top-right corner in Fig.~\ref{fig:oxygen_related}l).
Mineralization and nitrification is lower in the uniform-sensitivity case in Jan-Apr (Fig.~\ref{fig:oxygen_related}).

The oxygen produced by phytoplankton does not closely follow temperature changes, yet has higher rates in summer, rapidly increasing in Jul and decreasing in Aug (Fig.~\ref{fig:oxygen_related}m).
The relative difference between the uniform and reference case ranges from 0 to 35\%  (Fig.~\ref{fig:oxygen_related}n).
The relative between cases reduces from 0 to 14 \degree\ and decreases with increasing temperature, but showing higher differences at 14 \degree\ than the expected by the trend observed at lower temperatures (high-density spot in the top-right corner in Fig.~\ref{fig:oxygen_related}o).
The oxygen consumed by respiration follows a similar pattern to that produced by phytoplankton, with similar observations about its variability with temperature (Fig.~\ref{fig:oxygen_related}p-r).
During summer, all processes produced or consuming oxygen are higher in the uniform-sensitivity case (i.e., $E<0$ in Jul--Oct in center column in Fig.~\ref{fig:oxygen_related}).

\subsection{Oxygen flux}

\begin{figure*}[h] 
 \centering
 \includegraphics[width=0.99\textwidth]{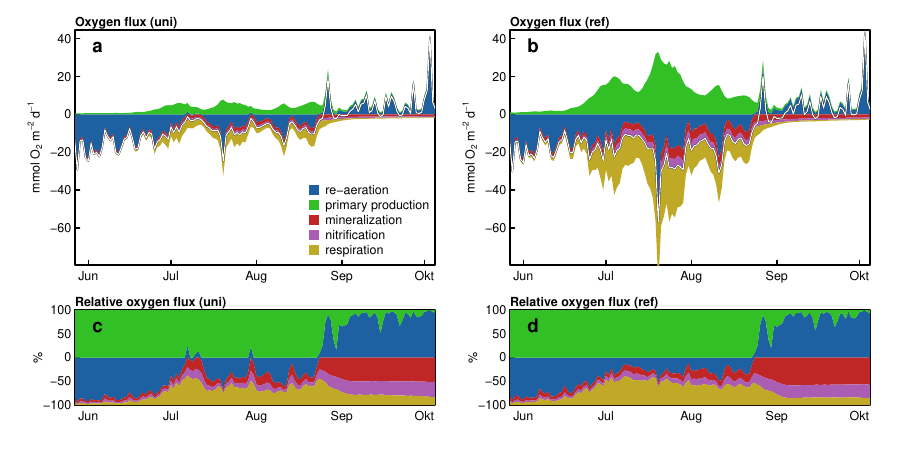} 
 \caption{\textbf{Absolute and relative (top and bottom, respectively) oxygen fluxes in uniform sensitivity and referent cases (left and right, respectively) for summer.} 
 The white line in panels a and b is the sum of all processes.
 }
 \label{fig:oxygen_flux}
\end{figure*}

The absolute oxygen flux leading oxygen dynamics shows large differences in summer, depending on the assumed temperature sensitivities  (Fig.~\ref{fig:oxygen_flux}).
The absolute oxygen flux with uniform temperature sensitivities is 2-to-5 fold lower than the reference case (Fig.~\ref{fig:oxygen_flux}a and b).
The magnitude of strong deoxygenation events is missing in the uniform sensitive case in Jul and Aug.
Uniform and reference cases both show similar relative oxygen flux, except for July (Fig.~\ref{fig:oxygen_flux}c and d).
Photosynthesis is the primary source of oxygen, re-aeration is an oxygen sink, and respiration is the dominant sink for oxygen.
Mineralization and nitrification became more important from late August, when the role of re-aeration became a source.

\section{Discussion}

Our results show a strong effect of temperature sensitivity in the ecological variables and seasonal variability of the processes explaining oxygen dynamics, both in relative and absolute terms.
In OxyPOM we consider well‑established temperature dependencies for several key processes: re-aeration (in the Sc number, after \citet{Wanninkhof2014}), maximum growth rates of phytoplankton (Eq.~\ref{eq:rho}) \citep{Coles2000, Sherman2016, Anderson2021}, optimal light responses (Eq.~\ref{eq:light-temp}) \citep{Coles2000, Edwards2016}, grazing and respiration (Eqs.~\ref{eq:m} and \ref{eq:r}) \citep{Verity2002, Scharfe2009, Ferreira2022}, biochemical rates (Eqs.~\ref{eq:MX} and \ref{eq:gamma}) \citep{Krishnan2014, Velthuis2022}, and viral infections (Eqs.~\ref{eq:n} and \ref{eq:B}) \citep{Padhy1977, DePaepe2006}. 
These are incorporated through a $\Qten$ rule (Eq.~\ref{eq:Q10}) with specific sensitivities for each process.
The observed $\Qten$ values span a wide range across different processes (Table~\ref{table:q10}) and thus constitute a major source of uncertainty and variability in model applications, significantly affecting model responses and results interpretations \citep[e.g.,][]{Maar2011, Olonscheck2013}.

The wide range of $\Qten$ highlights two key points about our modelling approach: 
First, both the uniform and reference cases are equally realistic---there is not a ``good'' and a ``bad'' parametrization---because we employ $\Qten$ values that in either case fall within the range observed in nature. 
Second, capturing this variability requires a model that explicitly incorporates temperature‑sensitivity parameters. 
Depending on the intended application, one can either use a single, unified parameter for all processes or assign distinct, process‑specific parameters \citep[e.g.,][]{Soetaert2001,LeQuere2016}. 
OxyPOM is capable of supporting either strategy.

\subsection{Temperature‑dependent biases in carbon, nutrient, and virus pools}
Temperature sensitivities modify the entire ecosystem response and are not limited to effects in oxygen concentration.
We observed this across most of the simulated year, the uniform‑sensitivity runs produced modest overestimates of dissolved oxygen in the bottom water column during Jul--Oct. 
This excess in oxygen coincides with the underestimation of carbon‑related pools (phytoplankton, DOC, and total POC) and overestimation of nutrients (ammonia, nitrate, and phosphate) in the same spatial and temporal window (Fig.~\ref{fig:main_vars}). 
This pattern suggests that a uniform temperature response may dampen the coupling between primary production, organic matter mineralization, and subsequent nutrient regeneration.
This coupling is critical for understanding experiments with temperature manipulation \citep{Biermann2014}.

Surprisingly, increasing the $Q_{10,\text{Phy}}$ from $1.7$ in the reference case to $2.5$ in the uniform case did not increase the phytoplankton concentration in summer (Fig.~\ref{fig:main_vars}f) and instead reduced it.
Previous modelling efforts have highlighted the importance of temperature‑dependent phytoplankton growth rates for reproducing seasonal bloom dynamics \citep[e.g.,][]{Maar2011, Behrenfeld2013}. 
Our results agree with these observations, since a uniform-temperature sensitivity not only suppresses the magnitude of the bloom (the reference case exhibits a higher phytoplankton peak concentrations) but also propagates downstream to affect the balance of oxygen and nutrients. 
The underestimation of phytoplankton carbon in the uniform case reduces grazing, respiration and mineralization, thereby limiting oxygen consumption and nutrient release. 

Despite having an explicit temperature sensitivity, pathogenesis, here in form of viral infections, had only a small response to a uniform parameterization (Fig.~\ref{fig:oxygen_related}a-c).
Virus-mediated mortality of diatoms showed only minor differences between the two configurations, with the largest deviations occurring in Jul--Oct and occasional peaks in April. 
Relative differences become apparent only within a narrow temperature band (10--14 \degree), while remaining near zero elsewhere. 
This apparent limited sensitivity aligns with experimental observations that viral infection rates often display weaker temperature dependence than host growth \citep{Middelboe2001}. 
The reduced temperature‑driven variation observed here could nevertheless influence cascade effects on carbon export and nutrient cycling.

\subsection{Compensatory effects of temperature sensitivity in oxygen budget}

Denitrification, mineralization, and nitrification all exhibited clear seasonal trends that follow temperature (Fig.~\ref{fig:oxygen_related}d-l).
Mineral denitrification and mineralization share a common $\Qten$ value in our model configuration.
Notably, the parameterization with uniform temperature sensitivity tended to underestimate process rates during the warmest months, leading to a 2‑to‑5‑fold reduction in absolute oxygen flux compared with the reference case. 
This discrepancy shows the importance of assigning process‑specific $\Qten$ values: a single temperature coefficient cannot capture the divergent thermal optima of microbial pathways, e.g., denitrifying bacteria typically have higher temperature optima than nitrifying bacteria \citep{Sherman2016, Mundim2020, Velthuis2022}.

Oxygen production by phytoplankton and consumption by respiration displayed a more complex relationship with temperature.
Additional factors, such as light and nutrient concentrations may limit photosynthesis rates, regardless of the temperature response \citep{Edwards2016, Holzwarth2018a}.
While temperature is the most reported predictor of respiration, organic carbon supply (from primary production) shows similar importance, suggesting their complex interactive relationship \citep{Wikner2023}.
Photosynthesis and respiration peaked in summer, yet their relative differences were just weakly temperature‑dependent. 
This suggests that factors beyond temperature, such as light availability, nutrient limitation, grazing, and viral infections may play a dominant role in governing net community metabolism during the bloom period \citep{Edwards2016,Wirtz2019}.

\subsection{Temperature sensitivities: a source of uncertainty}

Temperature sensitivities reported in the literature are highly variable:
For example, the estimated $\Qten$ values for photosynthesis depend on many factors and degrees of aggregation. 
For specific species, \citet{Coles2000} report $\Qten$ for photosynthesis rates of cyanobacteria and diatoms of 1.79--2.67 in the range of 2--25 \degree, while \citet{Ferreira2022} report values as high of 3.16 for dinoflagellates.
In contrast, community aggregated $\Qten$ values for phytoplankton growth rate are highly variable, e.g., 2.2--2.5 \citep{Verity2002} and 1--7 \citep{Robinson1993}, the later for Antarctic communities.
Meta-analysis of experimental data show taxa and habitat dependent $\Qten$ ranging from 1.42 for marine coccolithophores to 4.9 for marine cyanobacteria \citep{Anderson2021}, while field experiment estimations of global sampling are more conservative with a mean of $1.47 \pm 0.08$ \citep{Sherman2016}.

Carbon loss rates show similar variability:
$\Qten$ values for respiration rates of dinoflagellates and protozooplankton are 1.1--1.85 for temperatures in the range 16--22 \degree\ \citep{Ferreira2022}.
However, other planktivorous groups show higher values, for example 2.15--4.6 for copepods \citep{Hirche1987} and 3.0 for cod larvae \citep{Peck2008}.
Phytoplankton respiration $\Qten$ at the community level are in the range 1.45--5.59 \citep{Robinson1993}, and in temperatures approaching zero growth above 10 are possible but uncommon \citep{Pomeroy2001a}.
Values lower than one, indicating that respiration rates reduce with temperature, have been also observed \citep{Wikner2023}.
\citet{Verity2002} report size-dependent $\Qten$ for grazing rates ranging 2.2--2.9, yet higher ranges have been observed and used in modelling 2.2--4.8, the higher values corresponding to copepods \citep{Maar2011}

Turnover rates for organic carbon and nitrogen also have wide range. 
\citet{Hansen2014} report $\Qten$ for mineralization ranging 1.5--5.0.
Lower $\Qten$ values, approximately 1, are observed for DOC mineralization \citep{Johannsson2025}.
$\Qten$ for nitrification rates ranging 0.2--2.9  \citep{Krishnan2014} and  0.8--1.3 in Arctic has been observed \citep{Baer2014}.
Other nitrogen related processes, such as denitrification range a little higher 2.0--2.5 \citep{Krishnan2014} and  $2.3 \pm 4.7$ \citep{Velthuis2022}. 

Temperature dependence of the optimal light intensity for photosynthesis is not often included in models, however $\Qten$ values ranging 1.62--1.85 have been calculated experimentally \cite{Edwards2016}.
Values in this range can be estimated for low-temperature adapted species \citep[1.8 calculated from][]{Tilzer1986}, and for cyanobacteria and diatoms \citep[1.5 calculated from][]{Coles2000}.

As including virus in plankton models is still uncommon, available parameterizations for temperature are uncommon.
Burst size---the amount of virus copies produced by infected host---show a strong variability with temperature: \citet{Padhy1977} reports smaller burst size with increasing temperature, indicating $\Qten<1$, while other report larger burst size with increasing temperature \citep[e.g.,][]{Woody1995, Nagasaki2003, Maat2017}. 
These variations might be related to the lytic/lysogenic virus cycle \citep{Parada2006}.
In model applications, burst size is often kept constant \citep[e.g.,][]{Krishna2024}.
Virus life-time--related to inactivation rates--have been observed in the range 1.3--2.7 \citep[estimated from activation energies reported by][]{Yap2021a}.

\subsection{Room for further development}

Despite its oxygen-centered philosophy, OxyPOM neglects several biogeochemical processes involved---either directly or indirectly---in oxygen dynamics.
For example, mineralization and nitrification are the only biochemical pathways of oxygen consumption.
Other biochemical redox reactions, however, such as sulphur, iron, and manganese kinetics, are necessary to model oxygen consumption, especially in hypoxic and anoxic waters, and have been partially considered in other models, e.g., ERGOM, BROM, and AQUATOX \citep{Yakushev2017, Neumann2022, Park2002}. 

Additional limitations appear from the simplified lower-trophic levels.
For example, OxyPOM prescribes phytoplankton with a fixed stoichiometry, assuming a nutrient-replete system which may not be adequate for all ecosystems \citep{Neumann2022}.
Additionally, a single temperature-dependent mortality rate affects all trophic interactions, when in reality different predators may present different temperature niches, thus different temperature-dependent factors \citep{LeQuere2016, Anderson2021}.

Temperature dependence can have responses different than the $\Qten$  rule included in OxyPOM, e.g., linear temperature response for growth rates as suggested by \citet{Montagnes2003}.
Moreover, temperature dependencies can be weakened based on observed acclimation or even reverted in high temperature.
For example, \citet{Menden-Deuer2018} suggest that time scale-dependent temperature sensitivities have been observed for photosynthesis and grazing, and \citet{Edwards2016} calculate $\Qten$ values specific for photoinhibition.
Furthermore, temperature sensitivities have been weakened below to established ranges based on model re-calibration, for example $\Qten = 1.1049$ for zooplankton grazing \citep{Olonscheck2013}.

Other important aspects of oxygen budget, such as sediment and benthic process \citep{Yakushev2013, Salk2016}, are adjacent to pelagic oxygen dynamics and not explicitly included.
These limitations can be overcome by coupling OxyPOM with other models via the FABM interface.
Moreover, OxyPOM is an open-source project; modifications and extra functionality can be added as required by any user or developer. 

Despite these shortcomings, our model includes the most relevant processes that are required to explain oxygen dynamics in pelagic systems while offering  flexibility in the parameterization of temperature-dependent rates.
OxyPOM is thus a suitable model for studying oxygen dynamics in a warming world and during extreme physical--oceanographic events such as heatwaves \citep{Achterberg2025}.


\paragraph{Data availability.}
All results of our simulations can be reproduced using publicly available scripts at \url{https://github.com/ovgarol/elbe-oxygen} . 

\paragraph{Code availability.}
The source code for OxyPOM is archived in Zenodo \citep{Garcia-Oliva2025-oxypom}.
Custom scripts were used for simulation studies and data analyses, their code is publicly available at \url{https://codebase.helmholtz.cloud/dam-elbextreme/oxypom/-/tree/main/testcases/temperaure-sensitivity}.

\paragraph{Supplementary material.}
A model description following the emerging Purpose, Assumptions, Validation, Exploration (PAVE) model communication standard \citep{Bell2026} is available as supplementary material. 

\paragraph{Acknowledgments.} 
This study was made possible by grants No.~03F0954D of the German Federal Ministry of Research, Technology and Space (BMFTR) as part of the DAM mission ``mareXtreme'', project ``ElbeXtreme''; 
it was supported by the Helmholtz Association with their Innovation Pool for the Research Field Earth and Environment AGRIO: Effect of anthropogenic modifications and climate change on greenhouse gas emissions along the river-ocean continuum, and through the joint research program ``Changing Earth - Sustaining our Future''.
We thank the open source community for making available the tools used among them R, Linux \LaTeX\ and FABM.  The PAVE description and good software practices followed here were inspired by the Open Modeling Foundation (OMF).

\paragraph{Competing Interests Statement.}
We declare to have no competing interests.

\printcredits

\bibliographystyle{cas-model2-names}

\bibliography{references.bib}


\appendix
\numberwithin{equation}{section} 
\numberwithin{figure}{section} 
\numberwithin{table}{section} 

\section{Additional figures and tables}

\subsection{Full model parameterization}\label{sm:model-parameters}

Model parameters are show in in Table~\ref{table:parameters}.
Most of the parameters were taken from \citet{Holzwarth2018a} for the oxygen-related processes and from \citet{Wirtz2019} for the virus related processes.

\begin{table*}[b]
\centering
\caption{OxyPOM model parameters for the reference case, values, and units.}
\label{table:parameters}
\begin{tabular}{llll}
\hline
Parameter name & Symbol & Value & Units \\

\hline 

Stoichiometric ratio N:C phytoplankton & $a_\text{N}$ & 0.151 & mmol N mmol C\textsuperscript{-1} \\ 
Stoichiometric ratio P:C phytoplankton & $a_\text{P}$ & 0.0094 &  mmol P mmol C\textsuperscript{-1} \\ 
Stoichiometric ratio Si:C diatoms & $a_\text{Si}$ & 0.1413 &  mmol Si mmol C\textsuperscript{-1} \\ 
Silicate equilibrium concentration & $\textrm{Si}^*$ & 357.0 & mmol Si m\textsuperscript{-3} \\ 
Opal dissolution reaction rate constant & $k_\text{Si}$ & 3 $\times 10^{-6}$ & d\textsuperscript{-1}  \\ 

\hline 

Primary production rate (20\degree) & $\mu^*$ & 1.2 & d\textsuperscript{-1} \\ 
Mortality rate phytoplankton & $m^*_i$ & 0.5 & d\textsuperscript{-1} \\ 
Critical NH$_4$ concentration for uptake & $K_{\text{NH}_4}$ & 0.7 & mmol N m\textsuperscript{-3}  \\ 
Half-saturation N for production & $K_\text{N}$ & 0.36 & mmol N m\textsuperscript{-3}  \\ 
Half-saturation P for production & $K_\text{P}$ & 0.03 & mmol P m\textsuperscript{-3}  \\ 
Half-saturation Si for production & $K_\text{Si}$ & 1.0 & mmol Si m\textsuperscript{-3} \\ 
Winter grazing inhibition & $m_{\textrm{winter}}$ & 0.33 & -- \\ 

\hline 

Specific light extinction coefficient (Phyae) & $\epsilon_{\textrm{\scriptsize Phy}}$ & 0.0012 & m\textsuperscript{2} mmol C\textsuperscript{-1} \\ 
Specific light extinction coefficient (POC) & $\epsilon_{\textrm{\scriptsize POC}}$ & 0.0012 & m\textsuperscript{2} mmol C\textsuperscript{-1} \\ 
Specific light extinction coefficient (ISPM) & $\epsilon_{\textrm{\scriptsize ISPM}}$ & 0.03 &m\textsuperscript{2} g{-1} \\ 
Background light attenuation & $\zeta_0$ & 0.0 & m\textsuperscript{-1} \\ 
Optimal light intensity Phy$_1$ & $I^*_1$ & 25.0 & W m\textsuperscript{2}  \\ 
Optimal light intensity Phy$_2$ & $I^*_2$ & 30.0 &  W m\textsuperscript{2}  \\ 
Inorganic suspended particulate matter & ISPM & 47.0 & g \\ 

\hline 

Fraction released by autolysis & $f$ & 0.5 & -- \\
Mineralization rate POC$_L$ (20\degree) & $k_{\text{Min},L}$ & 0.1 & d\textsuperscript{-1} \\ 
Mineralization rate POC$_S$ (20\degree) & $k_{\text{Min},S}$ & 0.01 & d\textsuperscript{-1} \\ 
Mineralization rate DOC (20\degree) & $k_{\text{Min},D}$ & 0.03 & d\textsuperscript{-1} \\ 
Decomposition fraction from $L$ to $S$ & $\kappa_{L\rightarrow S}$ & 0.5 & -- \\ 
Decomposition fraction from $L$ to $D$ & $\kappa_{L\rightarrow D}$ & 0.5 & -- \\ 
Decomposition fraction from $S$ to $D$ & $\kappa_{S\rightarrow D}$ & 1.0 & -- \\ 
Half saturation of DO consumption in mineralization & $K'_{\text{DO}}$ & 63.0 &  mmol O\textsubscript{2} m\textsuperscript{-3} \\ 

\hline 

Maintenance respiration rate Phy$_1$ & $r^*_1$ & 0.036 & d\textsuperscript{-1}  \\ 
Maintenance respiration rate Phy$_2$ & $r^*_2$ & 0.045 & d\textsuperscript{-1}  \\ 
Respiration stoichiometric factor & $c$ & 0.7 & -- \\
Growth--respiration factor Phy & $\pi$ & 0.065 & -- \\ 

\hline 

Nitrification rate constant & $k_{\text{Nit}}$ & 28.6 & d\textsuperscript{-1} \\ 
Half-saturation of NH$_4$ in nitrification & $K_{\text{NH\textsubscript{4}}}$ & 36.0 & mmol N m\textsuperscript{-3}  \\ 
Half-saturation of DO in nitrification & $K_{\text{\DO}}^*$ & 31.3 & mmol O\textsubscript{2} m\textsuperscript{-3} \\ 
Half-saturation of NO$_3$ in denitrification & $K_{\text{NO\textsubscript{3}}}$ & 36.0 & mmol N m\textsuperscript{-3}  \\ 
Half saturation of DO in denitrification & $K_{\text{\DO}}$ & 94.0 & mmol O\textsubscript{2} m\textsuperscript{-3} \\ 

\hline 

Settling velocity POC & $v_{\textrm{\scriptsize POC}}$ & -1.5 & m d\textsuperscript{-1}  \\
Settling velocity phytoplankton & $v_{\textrm{\scriptsize Phy}}$ & -0.1 &  m d\textsuperscript{-1} \\

\hline 


\hline 

Re-aeration geometry factor & $\alpha$ & 1.0 & -- \\ 

\hline 

Virus steepness (sensitivity) parameter & $S$ & 2.0 & -- \\ 
Virus replication rate & $G^*$ & 0.30 & -- \\ 
Virus inactivation rate & $B^*$ & 0.15 & -- \\ 
Virus host defense rate & $H_i^*$ & 0.2 & -- \\ 
Concentration of Phyae for host defense & $C$ & 1.0 & -- \\ 
Maximum virus concentration & $\textrm{Vir}^*$ & 2.0 & -- \\ 
Minimum virus concentration & $\textrm{Vir}_{\min}$ & 0.01 & -- \\ 
\hline
\end{tabular}
\end{table*}


\begin{figure*}[tb] 
 \centering
 \includegraphics[page=2,width=1.0\textwidth]{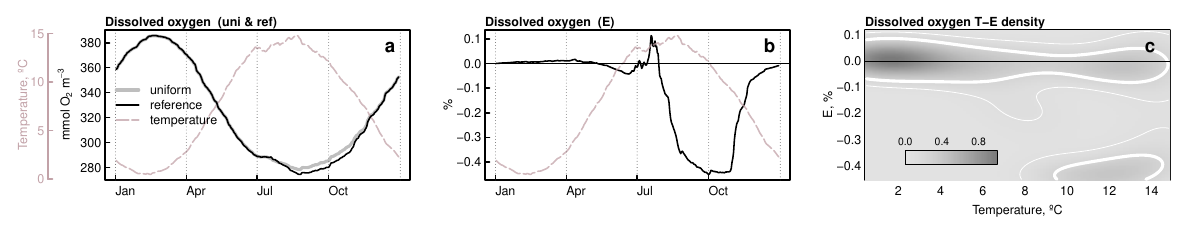} 
 \includegraphics[page=4,width=1.0\textwidth]{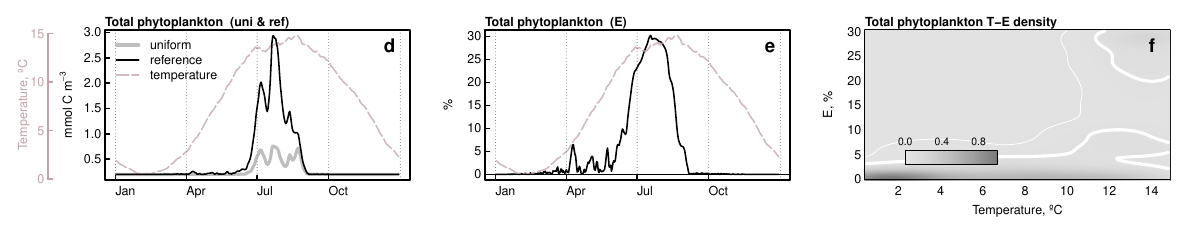} 
 \includegraphics[page=6,width=1.0\textwidth]{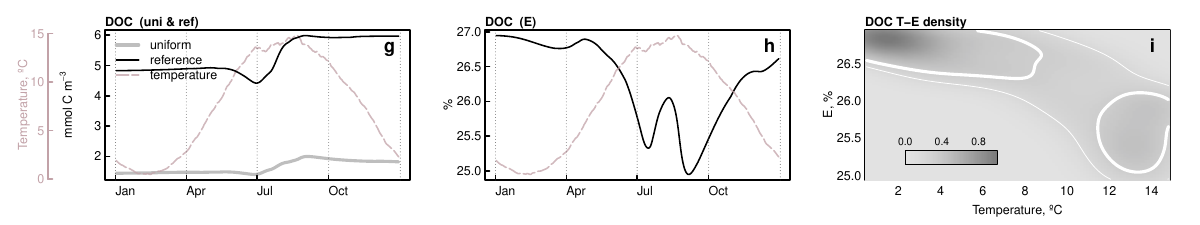} 
 \includegraphics[page=8,width=1.0\textwidth]{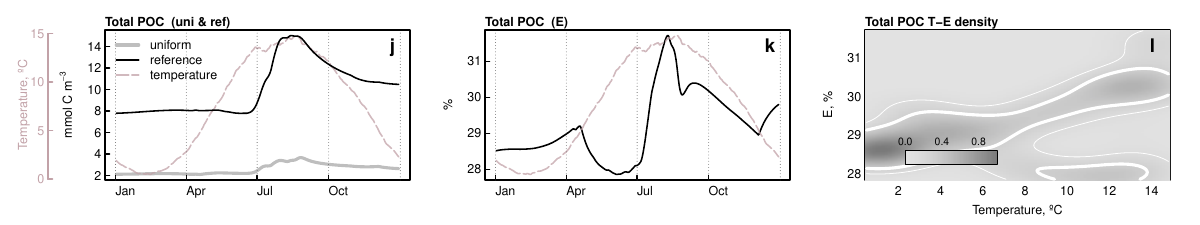} 
 \includegraphics[page=10,width=1.0\textwidth]{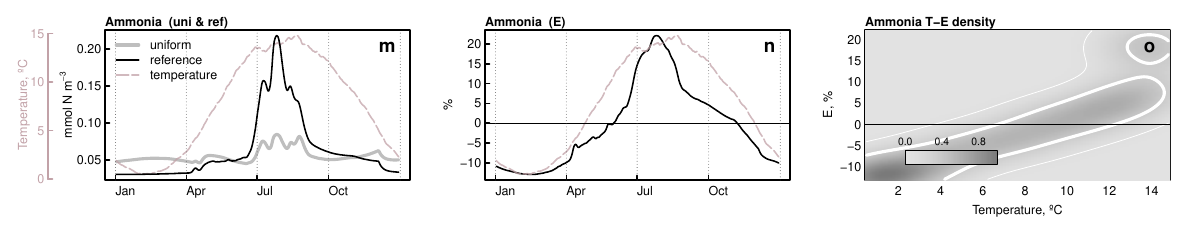} 
 \includegraphics[page=12,width=1.0\textwidth]{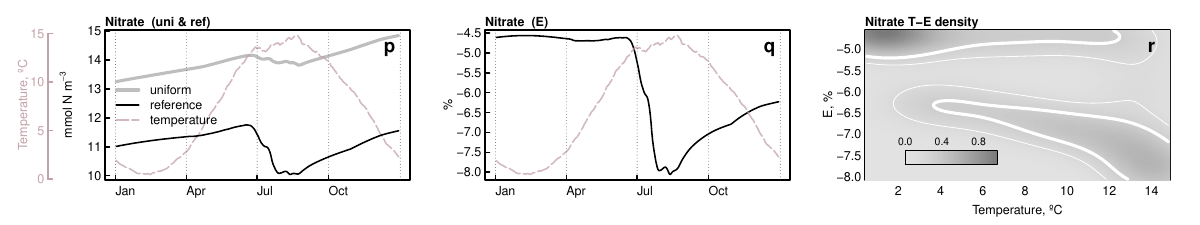} 
 \includegraphics[page=14,width=1.0\textwidth]{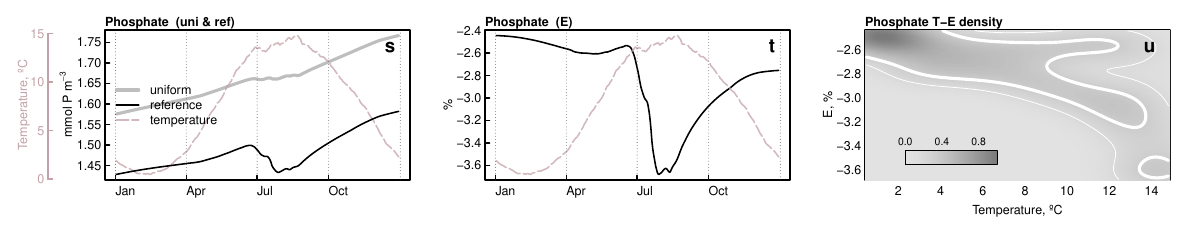} 
 \caption{\textbf{Spatial and temporal dynamics of key ecological variables in uniform-sensitive (uni) and reference (ref) cases, percent relative difference $E$, and $T$-$E$ density plot (left, center, and right columns, respectively).} 
 Dissolved oxygen (a-c), total phytoplankton (d-f), DOC (g-i), total POC (j-l), ammonia (m-o), nitrate (p-r), and phosphate (s-u).
 }
 \label{fig:main_vars2}
\end{figure*}

\begin{figure*}[tb] 
 \centering
\includegraphics[page=15,width=1.0\textwidth]{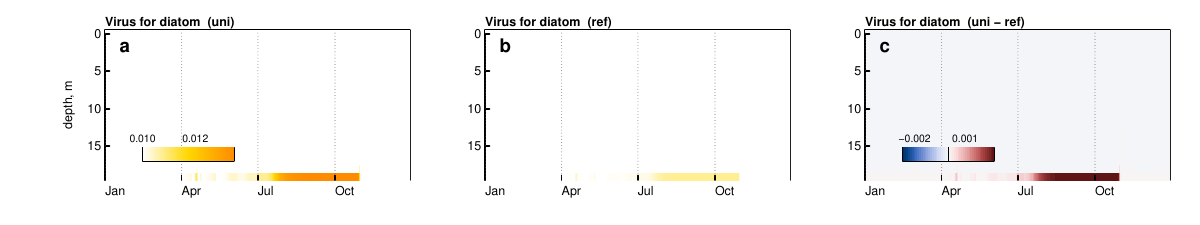} 
 \includegraphics[page=17,width=1.0\textwidth]{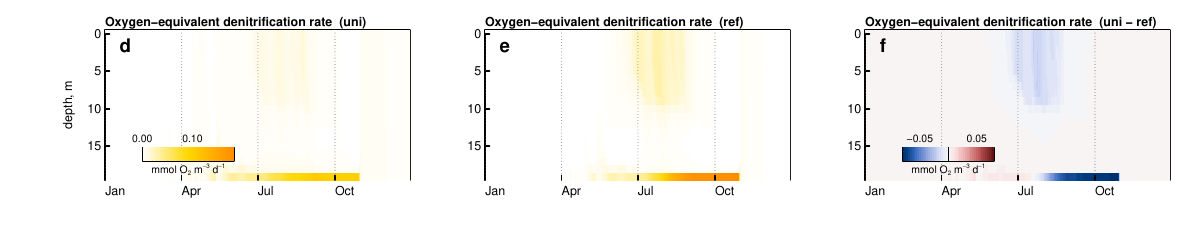} 
 \includegraphics[page=19,width=1.0\textwidth]{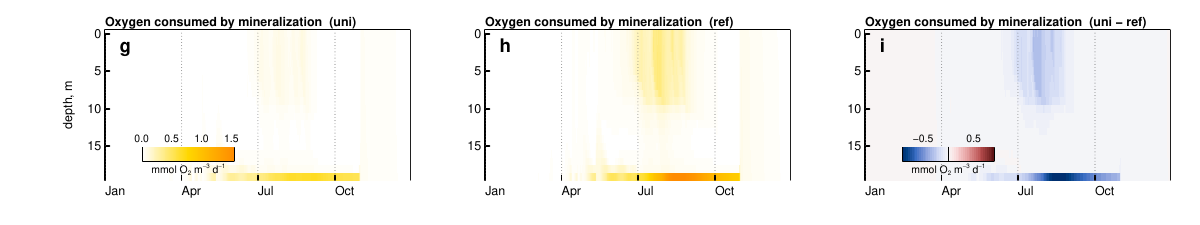} 
 \includegraphics[page=21,width=1.0\textwidth]{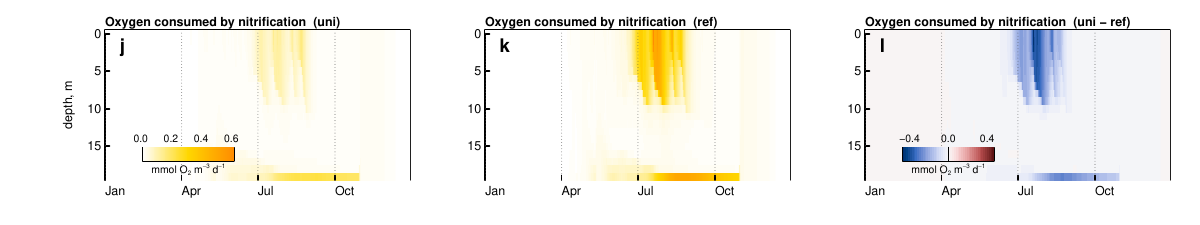} 
 \includegraphics[page=23,width=1.0\textwidth]{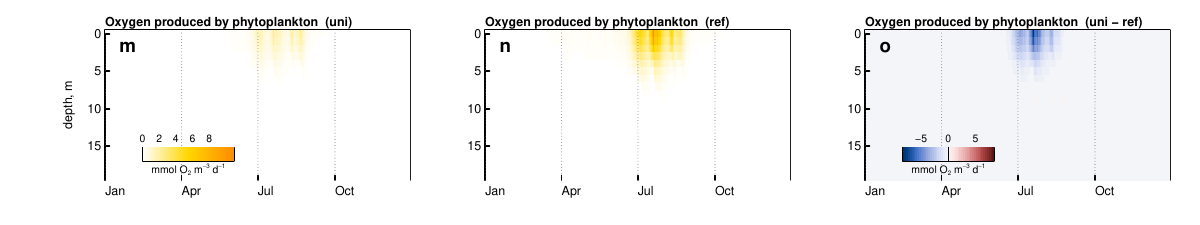} 
 \includegraphics[page=25,width=1.0\textwidth]{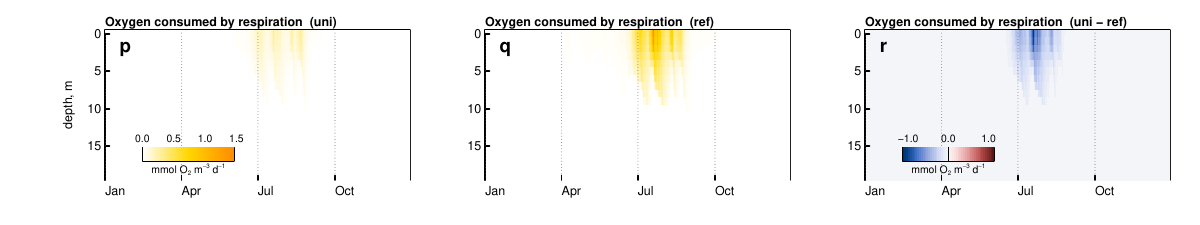} 
 \caption{\textbf{Vertical-averaged virus and oxygen-related process in uniform-sensitive (uni) and reference (ref) cases, and their difference (left, center, and right columns, respectively).} 
 Virus for diatom (a-c), denitrification (d-f), oxygen consumed by mineralization (g-i) and nitrification (j-l), produced by phytoplankton (m-o), and consumed by respiration (p-r).
 The pink dashed line shows the temperature.
 In the density plots (right column), white lines show the contour containing 75\% and 50\% of the values (thin and thick lines, respectively).
 }
 \label{fig:oxygen_related2}
\end{figure*}

\end{document}